\documentclass[secnumarabic,twocolumn,aps,showpacs,showkey,superscriptaddress,floatfix]{revtex4-1}

\usepackage{amsmath,amssymb,amsthm}

\usepackage{hyperref}
\usepackage{amsthm}
\usepackage{graphicx}
\usepackage{subfigure}
\usepackage{color}
\usepackage[normalem]{ulem}

\definecolor{violet}{rgb}{1.00,0.00,1.00}	
\definecolor{orange}{rgb}{1.00,0.50,0.00}	

\definecolor{turq}{rgb}{0.00,0.6,1.00}	

\newcommand{\alain}[1]{#1}
\newcommand{\john}[1]{#1}

\graphicspath{{./Figures/}}

\newcommand{\R}{\mathbb{R}}

\newcommand{\der}[2]{ \frac{\text{d} #1}{\text{d} #2} }  

\theoremstyle{plain}

\newtheorem{definition}{Definition}

\begin{document}

\title{Power-law statistics and universal scaling in the absence of criticality}

\author{Jonathan Touboul}%
\email[]{jonathan.touboul@college-de-france.fr}
\affiliation{The Mathematical Neuroscience Laboratory, CIRB / Coll\`ege de France (CNRS UMR 7241, INSERM U1050, UPMC ED 158, MEMOLIFE PSL*)}
\affiliation{MYCENAE Team, INRIA Paris}

\author{Alain Destexhe}%
\affiliation{Unit for Neurosciences, Information and Complexity (UNIC),
CNRS, Gif sur Yvette, France}
\affiliation{{The European Institute for Theoretical Neuroscience (EITN), Paris}}

\date{\today}%

\begin{abstract}


  Critical states are sometimes identified experimentally through
  power-law statistics or universal scaling functions. We show here
  that such features naturally emerge from networks in self-sustained
  irregular regimes away from criticality. In these regimes, statistical 
  physics theory of large interacting systems predict a regime where 
  the nodes have independent and identically distributed dynamics. 
  We thus investigated the statistics of a system in which units are 
  replaced by independent stochastic surrogates, and found the same power-law
  statistics, indicating that these are not sufficient to establish
  criticality.  We rather suggest that these are universal features of
  large-scale networks when considered macroscopically.  These results
  put caution on the interpretation of scaling laws found in nature.


\end{abstract}

\maketitle
\section{Introduction}

Power law statistics are ubiquitous in physics and life sciences. They
are observed in a wide range of complex phenomena arising in natural
systems, from sandpile avalanches to forest
fires~\cite{bak1997nature,jensen1998self}, earthquakes amplitude solar
flares, website frequentation as well as
economics~\cite{newman2005power}. These distributions reveal unusual
properties: the observed quantity has no typical scale, and is not
well characterized by its mean (when it exists): it is distributed
over orders of magnitudes and large deviations are not exceptionally
rare, in the sense that extreme events are far more likely than they
would be, for instance, in a Gaussian distribution. These singular
properties, combined with their ubiquity in nature, has attracted wide
attention in applied science.

{A number of theories were proposed in order to account for the
  presence of such power-law distributions. Some theories use an
  analogy with statistical physics systems and associate the presence
  of power-law scalings as the hallmark that the system could be
  operating at a phase transition. These theories associate power-law
  to a so-called notion of \emph{criticality}: classically, in
  statistical physics, \emph{critical phenomena} are the behaviors
  occurring in systems in association with second order phase
  transitions. These are largely thought to be universal although
  no proof was provided yet. Indeed, a number of statistics are found
  in vastly distinct models, and particularly in the Ising
  model~\cite{kardar2007statistical} poised at its phase transition.
  The \emph{critical} regime thus occurs only at very specific
   parameters values.} At this regime the properties of the system 
   are particularly singular; in particular, a number of statistics
    are scale-invariant including for instance the size, duration 
    and shape of collective phenomena. The
seminal work of Bak, Tang and Wiesenfeld on the Abelian sandpile
model~\cite{bak1987self} largely popularized the hypothesis that
criticality may be the origin of power-laws observed in nature.
Indeed, despite the fact that parameters for criticality are very
rare, systems may {self-}organize naturally at criticality (at
the phase transition point) without requiring fine tuning by a
mechanism called self-organized criticality
(SOC)~\cite{bak1997nature}.  This remarkable theory has sometimes led
to the conclusion that natural systems were critical based on the
identification of power-law relationships in empirical data
(see~\cite{bak1987self,jensen1998self} for reviews).

While power laws, as well as phase transition (thus criticality) are
well identified in models, a number of authors have underlined the
importance of being cautious when claiming power-law behavior in
finite systems, questioning their relevance or
usefulness~\cite{stumpf2012critical,avnir1998geometry}.  In
particular, Stumpf and Porter in~\cite{stumpf2012critical} aptly noted
the importance to take a nuanced approach between theoretical and
empirical statistical support reporting a power-law, as theories arise
from infinite systems while and real systems and usual datasets are
finite.

In physics, several alternative theories were proposed to account for
the presence of power-laws (see e.g.~\cite{newman2005power} for a
review). In particular, it was noticed very early that a pure random
typewriter (a monkey sitting at a typewriter) would generate texts
with a power-law distribution of word
frequencies~\cite{miller1957some} identical as the one observed in
data. This work brilliantly showed that power-law may arise from
purely stochastic mechanisms, evidencing that some power-laws
distributions may not reflect deep structures in the data.
Li~\cite{li1992random} formalized Miller's theory, highlighting the
fact that combinations of exponentials naturally yield power-law
relationships.
Bouchaud and others~\cite{bouchaud1995more,jan1999fossil} noted that
inverse of regularly distributed quantities may show power-law
distributions. Newman and colleagues showed that random walks generate
several statistics scaling as power-laws~\cite{newman2005power}, Yule
introduced a process with broad applications, particularly in
evolution, naturally associated to power-law distributions, and
Takayasu and
collaborators~\cite{takayasu1991statistical,takayasu1988power,takayasu1989steady}
showed that systems with aggregation and injection naturally generate
clusters of size scaling as a power-law with exponent $-3/2$. In
neuroscience, Benayoun, Wallace and Cowan have shown that neuronal
networks models in a regime of balance of excitation and inhibition
also provide power-law scalings of
avalanches~\cite{benayoun2010avalanches}. All these mechanisms are
independent of any phase transition and arise away from criticality
from a particular way of considering a random process.

The hypothesis that networks of the brain operate at criticality was
introduced a decade ago with the development of recording techniques
of local populations of cells and the analysis of specific events
corresponding to collective bursts of activity separated by periods of
silence.  The first empirical evidence that neuronal avalanches may
show power-law distributions of duration or size was derived from the
analysis of neuronal cultures \emph{in vitro}
activity~\cite{beggs2003neuronal}. Based on an analogy with the
sandpile model, these bursts were seen as ``neuronal avalanches'', and
were apparently distributed as a power-law with a slope close to
$-3/2$, consistent with the distribution of sandpile avalanches. These
\emph{in vitro} findings were based on indirect evidences of spiking
derived from local field potentials, extracellular signals associated
with the summation of postsynaptic potentials (bursts produce negative
peaks in the LFP signals) and affected by a number of events unrelated
with spiking activity.  Similar LFP statistics were later found
\emph{in vivo} in the awake monkey~\cite{petermann2009spontaneous} and
in the anesthetized rat~\cite{hahn2010neuronal}.  These empirical
evidences were used to draw strong conclusions on neural coding: the
presence such power-laws would ensure maximized information capacity,
dynamic range and information
transmission~\cite{shew2013functional,chialvo2006psychophysics}.
However, the method of analyzing the amplitude of negative LFP peaks
was shown to produce spurious power laws
scalings~\cite{touboul2010can} regardless of the spike activity of
cells. Indeed, identical scalings were found in surrogate data,
positive LFP peaks (that are independent of spiking activity), and
also arise in elementary purely stochastic signals, \john{such as 
excursions of Ornstein-Uhlenbeck processes through thresholds away 
from the mean, or in one-dimensional random walks~\cite{colaiori2004average,
baldassarri2003average}: both duration and time of excursions show 
power-law statistics, and display shape invariance. 
It was further shown in that both in data and surrogate models, }
statistical significance of these power-laws of \john{LFP peak} was poor, and depended on
the threshold chosen. In~\cite{dehghani2012avalanche}, Dehghani and
collaborators have made a statistical analysis combining
multielectrode in vivo recordings from the cerebral cortex of cat,
monkey and human, and did not confirm the presence of power-laws. The
data rather showed an optimal fit with two independent exponential
processes.

The poor statistical significance of LFP avalanche analysis and the
ambiguous results it yields has motivated an in-depth exploration of
\emph{in vitro} spiking data of cultures of
neurons~\cite{friedman2012universal}. In this remarkable work, the
authors used multi-unit data from in vitro cultures and addressed a
number of properties of critical systems reported by Sethna \emph{et
  al} unified theory of criticality of statistical
systems~\cite{sethna2001crackling}. Of crucial importance in this
theory are power-law scalings of specific events and the relationship
between the different scaling exponents. Friedman and
collaborators~\cite{friedman2012universal} revealed that the data was
consistent with Sethna et al theory, since power-law scalings of both
avalanche size and duration were reported, with slopes consistent with
the critical exponents of $-3/2$ and $-2$ respectively, but also the
existence of a universal mean temporal profiles of firing events
collapsing under specific scaling onto a single universal scaling
function, thereby providing more substantial analogy between this {\it
  in vitro} system with statistical physics models at criticality.

In the present paper, we show that these observations can arise
naturally in neuronal systems that are not at criticality. {We
also provide a theoretical explanation for this, as well as 
analytic access to some of the relevant properties of such systems.}

\section{Avalanches in spiking network models}\label{sec:aval}

\begin{figure*}[h]
	\centering
		\includegraphics[width=.7\textwidth]{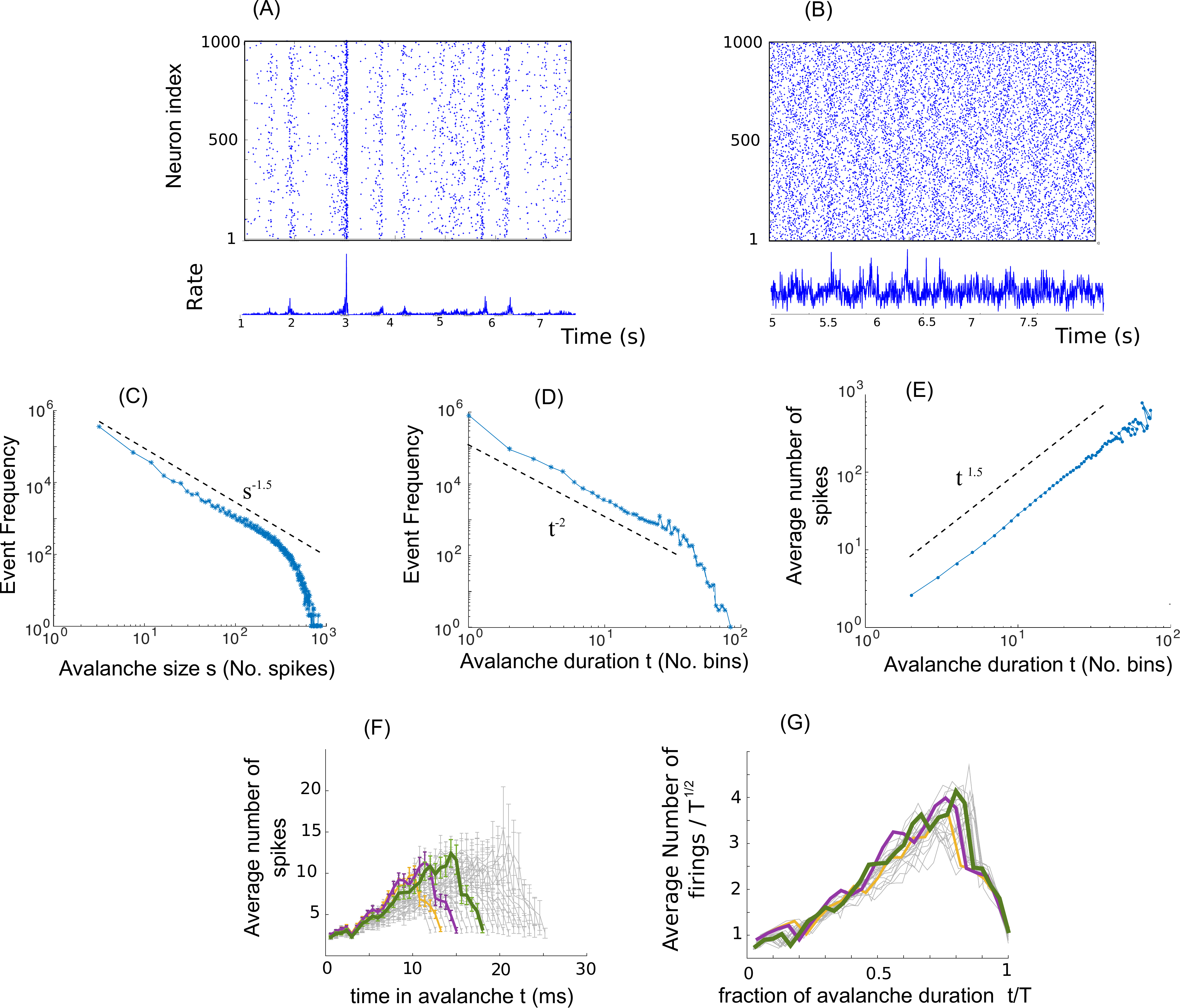}

                \caption{Avalanche spike statistics in the sparsely
                  connected networks~\cite{brunel2000dynamics} with
                  $N=5\,000$ neurons in the SI and AI states. (A)
                  Raster plot of the SI together with the firing rate
                  (below). (B) Raster plot in the AI state: spiking is
                  asynchronous and faster (notice that the time window
                  is shorter compared to A for legibility): no silence
                  period arises. {Parameters as
                    in~\cite[Fig.2D]{brunel2000dynamics} with $g=4$
                    (excitation-dominated regime) and
                    $\nu_{ext}/\nu_{thresh}=1.1$. }. In the SI states,
                  (C) avalanche size (number of spikes) and (D) avalanche 
                  duration (number of bins) scale
                  as power law (dashed line is the maximum likelihood
                  fit), and averaged avalanche size scales as a power
                  law with the duration avalanche duration according
                  to the universal scaling
                  law~\cite{sethna2001crackling} (E). Average
                  avalanche shapes collapse onto the same curve (F,G)
                  very accurately. {Avalanche shapes from 20 to
                    40 bins are plotted in gray, and 3 of the same
                    size as in~\cite{friedman2012universal} are
                    colored and with distinct thicknesses. 
                    Parameters as in~\cite[Fig.
                    2D]{brunel2000dynamics} with $N=1000$, $g=5$
                    (inhibition-dominated regime) and
                    $\nu_{ext}/\nu_{thresh}=0.9$.}  }

	\label{fig:Brunel}
\end{figure*}

We start by investigating the avalanche distributions generated by 
the classicalmodel of spiking neuronal network with
excitatory and inhibitory connections introduced by Nicolas Brunel
in~\cite{brunel2000dynamics}. This model describes the interaction of $n$
neurons described through their voltage $(v_i)_{i=1\cdots n}$ that decays to reversal potential in the
absence of input, receives external input and spikes from
interconnected cells, and fire action potentials when the voltage
exceeds a threshold $\theta$. In detail, the voltage of neuron $i$
satisfies the equation:
\begin{equation}\label{eq:Brunel}
\tau \der{v_i}{t} = -v_i + R\tau \sum_{j=1}^n J_{ij}\sum_{k\geq 0} \delta(t-t_j^k-D)
\end{equation}
while $v_i\leq \theta$, and where $\tau$ denotes the time constant of the membrane and $R$ its resistivity.
The input received by the neuron are Dirac $\delta$s. Neuron $i$ receives the $k$th impulse of neuron $j$, 
emitted at time $t_j^k$, after a delay denoted $D$ and assumed constant, which 
alters its membrane potential of a quantity proportional to $J_{ij}$. Brunel's model assumes that these
coefficients are zero except for a fixed number of cells randomly
chosen in the excitatory and inhibitory populations, for which the
coefficient $J_{ij}$ have fixed values $J$ and $-gJ$ respectively
(see~\cite{brunel2000dynamics} for details). This model is particularly
interesting very for its versatility and ability and to produce diverse
spiking patterns. One can classify the
regimes of activity in terms of levels of synchrony and regularity,
and different regimes emerge as a function of the relative levels of
excitation and inhibition and the input, that can be identified
through the computation of precise bifurcation
curves~\cite{brunel2000dynamics}. The thus obtained regimes are 
termed \emph{activity states}, to distinguish these from the statistical mechanics
notion of \emph{phase}: the regimes are here separated by bifurcations occurring 
in the mean-field limit of the system.
Of special interest are the {\it Asynchronous Irregular} (AI) states, in which neurons fire in a Poisson-like manner, with no period of silence (hence no avalanche). This activity is evocative of the spike trains observed experimentally in awake animals. Such sparsely-connected networks can also display periods of collective activity of broadly distributed duration interspersed by periods of silence, called {\it Synchronous Irregular} (SI) regimes, known to reproduce the qualitative features of spiking in anesthetized animals or neuronal cultures.  Although partially ordered and partially disordered, SI
regimes occur for a wide range of parameter values (all in inhibition-dominated regimes)~\cite{brunel2000dynamics} and are not very sensitive to modifications of biophysical parameters. The SI regime is not at a transition in the activity regime; within this region, chaotic activity takes place. Indeed, inhibition dominates the excitation, thus when the activity spreads throughout the network, it triggers massive inhibition that naturally silences the network. This is what we observe in simulations of the Brunel model (Fig.~\ref{fig:Brunel})\footnote{{We used the algorithm freely available on ModelDB}.}. We investigate the statistics of spike units in both cases (see Fig.~\ref{fig:Brunel}).

First, in the AI state, the sustained and irregular firing does not
leave room for repeated periods of quiescence at this network scale,
preventing the definition of avalanches (see raster plot in
Fig.~\ref{fig:Brunel}(B)) \footnote{Sub-sampling of the network may
  lead to find periods of quiescence, whose statistics will depend on
  the network size.  However, these are not robust statistical
  quantities to define a critical regime.}.
This sharply contrasts
with the SI state, in which we can define avalanches which
  display the same statistics assumed to reveal criticality in
cultures~\cite{friedman2012universal}. Fig~\ref{fig:Brunel}(A)
represents the raster plot with typical avalanches taking place.
Strikingly, both avalanche duration and avalanche size show excellent
fit to a power law, validated by Kolmogorov Smirnov test of maximum
likelihood estimator~\cite{clauset2009power}, and the exponents
found are consistent with those found in neuronal cultures~\cite{friedman2012universal}. In particular, we find that
the size $s$ of the avalanches (number of firings during one
avalanche) apparently scales as $s^{-\tau}$ with $\tau=1.42$
(Fig.~\ref{fig:Brunel}(C)), close to the theoretical value of $3/2$ (plotted on the figure for indication),
and to the neuronal culture scaling ($1.7$ reported
in~\cite{friedman2012universal}). Using the Kolmogorov-Smirnov test~\cite{clauset2009power}, we have tested the hypothesis that the data are distributed as a power-law, and validated the hypothesis (Kolmogov-Smirnov distance $0.0097$, p-value $p>0.99$. This is also the case of the distribution of the duration $t$
of the avalanches, found scaling as $t^{-\alpha}$ with $\alpha=2.11$
(Fig.~\ref{fig:Brunel}(D)), close to the theoretical value of $2$ for critical systems
and from the experimental value of $1.9$ found in neuronal
cultures~\cite{friedman2012universal}). The Kolmogorov-Smirnov distance with a pure power law is very low, evaluated at $0.031$, which leads to a high p-value $p=0.99$, validating clearly the hypothesis of a power-law distribution of avalanche durations. Similar to what was observed
experimentally, the fit is valid on two octaves, and drops down beyond\footnote{The Kolmogorov-Smirnov statistical tests is not affected by events
of very small frequency.}.
This exponential drop is classically related to subsampling effects. We validated the presence of finite-size cutoffs by varying the size of the system, and found indeed that as the network size increases, the fit with a power-law distribution is valid on larger time intervals and the cutoff only arises at later times (see Supplementary Figure~\ref{fig:subsampling}).  
Also consistent with neuronal data and crackling noise,
we found that the average avalanche size scales very clearly as a power of its duration, but with a positive
exponent $\gamma=1.50$, not consistent with the crackling noise relationship between exponents
\begin{equation}\label{eq:Sethna}
	\gamma=\frac{\alpha-1}{\tau-1}
\end{equation}
that predicts an exponent equal to $\gamma=2.64$, but however quantitatively consistent with the \emph{in vitro}
data of~\cite{friedman2012universal}.
We have also investigated the shape of the avalanches of different durations. We have found that, similarly
to critical systems or to \emph{in vitro} data, the avalanche shapes collapsed onto a universal scaling function
(Fig.~\ref{fig:Brunel}E) when time is rescaled to a unit interval and shape rescaled by $T^{\gamma-1}$ where $\gamma=1.5$
is the power law exponent of the average size.

These scalings are not specific to the particular choice of parameters
used in our simulations: we consistently find, in the whole range of
parameter values corresponding to the SI state of Brunel's model, and
in particular, away from any transitions, similar apparent power-law
scalings with similar scaling exponents (see supplementary
Figure~\ref{fig:BrunelAllSIs}).  \john{Moreover, we confirmed, in
  addition to the fact that this regime is away from all transitions
  between the different network activity, that the system is not at
  criticality, by showing that relaxation towards the SI regime after
  perturbation is fast (within milliseconds) within the region
  considered, although it does slow down close to the transition point
  (see supplementary Figure~\ref{fig:SlowDown}). } We conclude that
these statistics {are valid in a whole regime of the system where the
  activity is synchronous irregular, and \john{neither} at a
  transition of the model \john{nor in a regime consistent with the
    slow decay of perturbations associated to critical regimes.}
  Therefore, finding power-law statistics in neuronal avalanches with
  exponents $1.5$ and $2$ do not reveal that the system operates at
  criticality, but \alain{rather seems a property of} synchronous
  irregular states. }

{The SI states are prominent in neuronal activity, especially in
  anesthesia and neuronal cultures. It is precisely in these
  situations that power-law distributions of spike avalanches were
  reported experimentally~\cite{beggs2003neuronal,hahn2010neuronal}.
  This regime differs from the awake activity where neurons fire in an
  AI manner. In these regimes, power-laws and criticality were
  reported based on LFP recordings~\cite{petermann2009spontaneous}. We
  will come back to experimental evidences of power-laws in
  local-field potentials recordings of the activity in
  section~\ref{sec:LFP}. }

\section{Avalanches and Boltzmann molecular chaos property}

The observation of such scaling relationship in simple models of
neuronal networks away from criticality and for a broad range of parameter
values suggests that the observed
scaling is related to properties of the systems that are independent
of the notion of criticality and that may be relatively general. These may therefore
be related to the properties of the network activity, that we now describe
in more detail.

\subsection{Propagation of chaos in neural networks models}
The classical theory of the
thermodynamics of interacting particle systems states that in large
networks (such as those of the brain), the correlations between
neurons vanishes.  This is also known as Boltzmann molecular chaos
hypothesis (\emph{Sto\ss zahlansatz}) {in reference to the hypothesis
that the speed of distinct particles should be independent, key to the 
kinetic theory of gases of Ludwig Boltzmann~\cite{boltzmann1964lectures}. 
In mathematics, this property is called
\emph{propagation of chaos}, and is rigorously defined as follows:
\begin{definition}\label{def:Chaotic}
	For $(X_1^n,\cdots, X_n^n)$ a sequence of measures on $(\R^d)^n$. The sequence is said
	$X$-chaotic if for any $k\in \mathbb{N}$ and $i_{1},\cdots,i_k$ a set of indexes independent of $n$, we have
	$(X_{i_1}^n,\cdots,X_{i_k}^n)$ converge to $k$ independent copies of $X$ as $n\to \infty$.
\end{definition}
}
In our context, in the limit of large networks, neurons behave as independent jump processes with a
common rate, which is solution of an implicit equation. This
property is at the core of theoretical approaches to understand the
dynamics of large-scale networks~\cite{brunel2000dynamics,
renart2004mean,sompolinsky1988chaos,ostojic2014two}. In the case of Brunel's model, it is shown that since two neurons share a vanishing proportion of common input in the thermodynamic limit, allowing to conclude that the correlation of the fluctuating parts of the synaptic input of different neurons are negligible. This leads the authors to conclude that the spike trains of different neurons are independent point processes with an identical instantaneous firing rate $\nu(t)$ that is solution of a self-consistent equation (see~\cite[p. 186, first column]{brunel2000dynamics}). In that view, except in the case of constant firing rate (the asynchronous irregular state), neurons always show a certain degree of synchrony due to the correlations of the instantaneous firing rates of different neurons.

Mathematically,
several methods were developed for interacting particle systems and gases
(see e.g.~\cite{sznitman:89} for an excellent review). It is shown that generically,
systems of interacting agents, with sufficient regularity, show propagation of chaos.
All these results are in particular valid for neuronal networks, as was shown recently in
a number of distinct situations. Large $n$ limits and propagation of chaos was demonstrated for
large networks of integrate-and-fire neurons~\cite{robert2014dynamics}, firing-rate models with multiple populations~\cite{touboul2011noise},
 for conductance-based models even in the large time regime~\cite{CQ3}, and  was  shown
 to hold in realistic network models incorporating delays and
the spatial extension of the system~\cite{touboul2014propagation,touboul2014spatially}.
{Rigorous methods of convergence of particle systems show that the \emph{empirical measure}
of the system:
\[\hat{\mu}_n=\frac 1 n \sum_{j=1}^n\delta_{x_i}\]
converges in law towards a unique solution. A very powerful
and universal mathematical result demonstrated in~\cite[Lemma 3.1]{sznitman:84} ensures
that that the convergence of the empirical measure of a particle systems towards a unique measure
implies propagation of chaos. Several methods may be used to show convergence of the empirical 
measure, including in the case of neuroscience, coupling methods~\cite{touboul2014spatially},
compacity estimates~\cite{robert2014dynamics} or large deviations~\cite{cabana-touboul}.}

These results indicate that a universal form of activity emerges from neural networks, whereby neurons
are independent copies of the same process. Before investigating the avalanche statistics of
such regimes of activity, let us discuss the plausibility of the existence of these regimes
in neuronal data.

\subsection{Decorrelation in experimental data and models}
In natural environments, regularity of sensory input to the brain may create strong and long-range correlations in space and time~\cite{ruderman:94,ruderman-bialek:94,dong1995temporal,dan1996efficient}. It soon appeared that these correlations would be detrimental for the brain to encode sensory stimuli and detect changes efficiently~\cite{attneave:54, barlow:61}. Theoretical models of the visual system in particular have shown that decrease of redundancy by decorrelation was important for efficient encoding of natural images~\cite{laughlin:81,simoncelli2001natural,dimitrov1998spatial}. This was confirmed experimentally. In~\cite{peyrache2012spatiotemporal}, the authors used high density two-dimensional electrode array and found in particular a marked exponential decay of correlation of excitatory cells. An clear confirmation of decorrelation even for closeby cells receiving similar input was recently brought in a remarkable experimental work where chronically implanted multielectrode arrays were developed and implanted in the visual cortex of the macaque~\cite{ecker-berens-etal:10}. This protocol produced exquisite data allowing to show that even nearby neurons, generally thought to be strongly connected and to receive a substantial part of common input showed a very low level of correlation. Similar decorrelation results were reported in the rodent neocortex~\cite{renart-de-la-rocha-etal:10} with the same level of accuracy.

The origin of this decorrelation is still controversial and several assumptions were formulated, including the role played by adaptation ionic currents that could play central role in temporal decorrelation~\cite{wang2003adaptation},
negative correlations associated with the co-evolution of excitatory and inhibitory cells activity~\cite{renart-de-la-rocha-etal:10}, or sophisticated and robust mechanisms relying on neuronal nonlinearities and amplified by recurrent connectivities~\cite{wiechert2010mechanisms}, that was compatible with pattern decorrelation observed in the olfactory bulb of the zebra fish.

All these experimental findings confirm that regimes in which neurons are independent are plausible representations of neural networks activity. We now investigate the avalanche statistics of such networks.

\subsection{Statistics of networks in the molecular chaos regime}\label{sec:surrog}
Both the mathematical analysis of neuronal network models and fine analysis
of the structure of spike trains motivates the study of ensembles of neurons
that are independent but with common non-stationary statistics. The simplest
model made of one could think of is to consider a collection of
independent Poisson processes with identical time-dependent rates.

In that view, cells with constant firing rates resemble AI regimes.
To generate a stochastic surrogate of the SI regime, the common rate
of the cells should display non-periodically periods of silence.
An obvious choice would be to replay the rates extracted from the
SI state, and indeed such a surrogate
generated power-law statistics (not shown), but in this case we could
not rule out whether the power-law statistics are encoded in the rate
functions. To show that this is not the case, we generated a
surrogate independent of the rate functions, by using a common rate of
firing of the neurons given by the positive part $\rho_t^+$ of the
Ornstein-Uhlenbeck process:
\[\dot{\rho}_t=-\alpha \rho_t + \sigma \xi_t\]
with $(\xi_t)$ a Gaussian white noise. This choice is interesting in
that although periods of silence do not occur periodically, the duration
between two such silences have a finite mean. Actually, the distribution
of excursions shape and duration of the Ornstein-Uhlenbeck process are 
known in closed form~\cite{alili-patie-etal:05}. These distributions are, 
of course, not heavy tailed: they have exponential tails with exponent 
$\alpha$ which is the timescale of decay of the process. 

\begin{figure*}
	\centering
		\includegraphics[width=.6\textwidth]{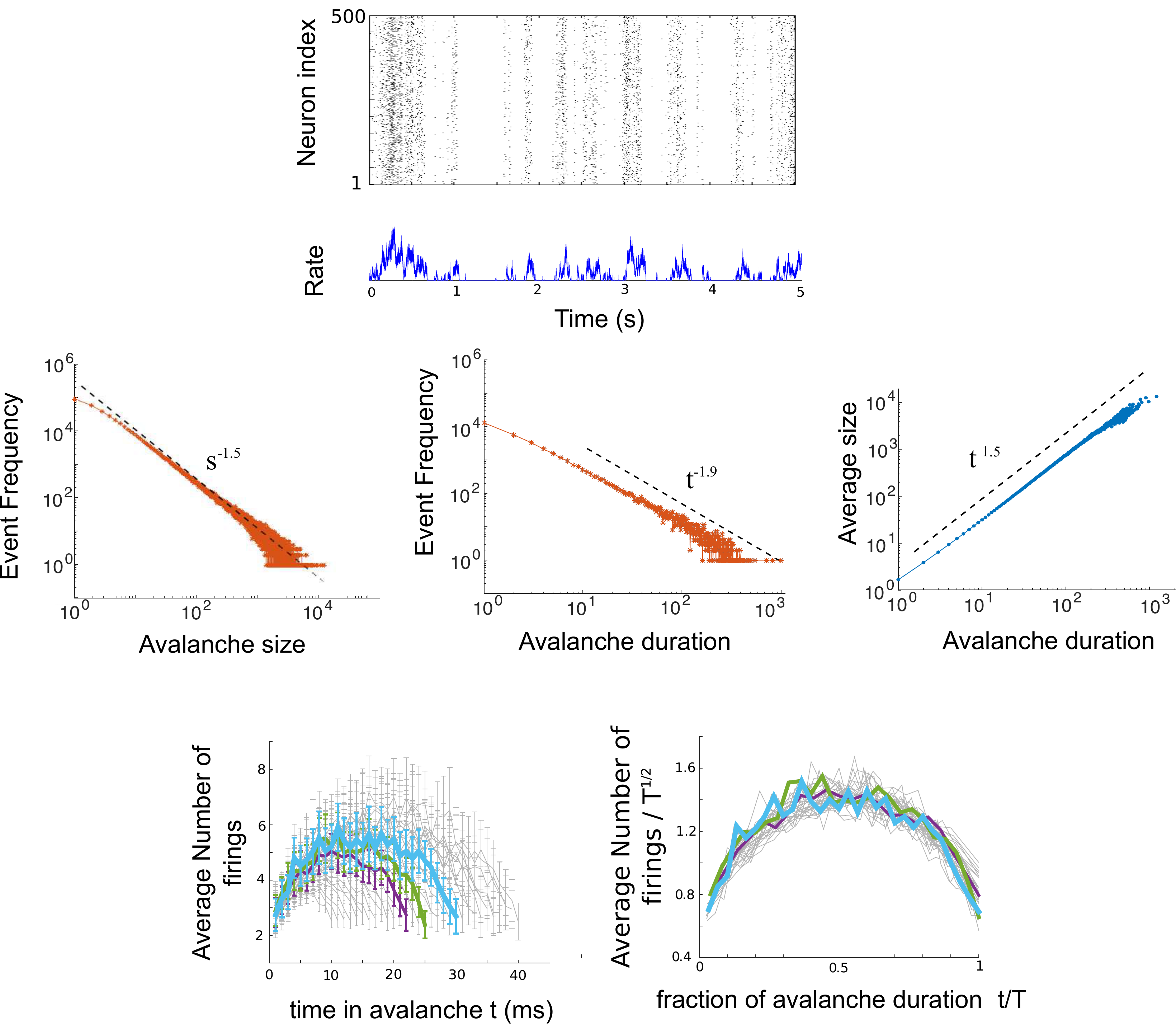}
                \caption{Avalanches statistics in the independent
                  Poisson model with Ornstein-Uhlenbeck firing rate
                  ($\alpha=\sigma=1$).  Apparent power-law scalings,
                  together with scale invariance of avalanches shapes.
                  {Avalanches of size 10 to 40 in gray, 3
                    specific trajectories highlighted}.}
	\label{fig:Poisson}
\end{figure*}

We investigated the collective statistics of $N=2\,000$ independent
realizations of Poisson processes with this rate. The resulting raster
plot is displayed in Fig.~\ref{fig:Poisson}(A).  While the firing is
an inhomogeneous Poisson process, macroscopic statistics show
power-law distributions for the size ($\tau=1.47$,
Fig.~\ref{fig:Poisson}(B)) and for the duration ($\alpha=1.9$,
Fig.~\ref{fig:Poisson}(C)), both statistically
significant~\cite{clauset2009power} and consistent with critical exponents. 
Again, a linear linear relationship between average avalanche size and duration 
is found, with a coefficient evaluated to $\gamma=1.4$ and explaining all the 
variance but $6\,10^{-4}$. Again, this coefficient is not consistent with the crackling 
relationship~\eqref{eq:Sethna}. Notwithstanding, we found that the averaged shape of 
avalanches of a given duration collapse on one universal curve when the amplitude is 
rescaled by the duration to the power $\gamma-1$ (Fig.~\ref{fig:Poisson}(D,E)).

Of course, these statistics are not related to the nature of the rate chosen.
We display in Supplementary Fig.~\ref{fig:ReflectedBrownian} for instance the
avalanche statistics of independent Poisson processes with rate given by the
positive part of a reflected Brownian, and we find exactly the same power-law
statistics of avalanche shapes and durations, as well as a very nice collapse
of avalanche shapes. 

\subsection{Analytical derivations in the stationary and slow rate regime}\label{sec:analytics}
{In this simple model, it is actually very simple to
indeed compute explicitly the distribution of avalanche duration for instance.
Disregarding the Poisson nature of the firings and the type of rate chosen, we
can indeed write down the probability for an avalanche of duration $t$ to occur
at a specific time. These distributions can be computed analytically in cases where
the spiking is described by a point process. In that case, denoting by $p(t)$ the probability for a neuron to spike in
the time interval $[t,t+\delta]$, the probability to observe an avalanche of size
$\tau$ starting at time $t^*$ in a collection of $n$ independent realizations is}
\[q(t^*)^n q(t^*+\tau+1)^n \prod_{t=1}^{\tau}(1-q(t^*+t)^n).\]
Assuming stationarity, the probability of finding an avalanche of duration $\tau$ is given by
\[\int_{[0,1]^{\tau+2}} q_0^n q_{\tau+1}^n \prod_{t=1}^{\tau}(1-q_t^n)d\rho_{\tau+2}(q_0,\cdots,q_{\tau+1})\]
where $\rho$ denotes the joint probability for the firing rate to have a specific sequence
of values $(q_0,\cdots,q_{\tau+1})$.

This formula remains quite complex. Assuming now that the rate is extremely
slow compared to the avalanches, one can simplify further the probability of 
$p_{\tau}^0$ of an avalanche of size $\tau$:
\[p_{\tau}^0=\int_{0}^1 q^{2n}(1-q^n)^\tau \rho_1(q) dq,\]
thus with a simple change of variable:
\[p_{\tau}^0=\frac 1 n \int_{0}^1 x^{1+\frac 1 n}(1-x)^\tau \rho_1(x^{\frac 1 n}) dx.\]
\john{As expected, when $n\to \infty$, the probability of having an avalanche 
of prescribed, finite duration goes to zero as $1/n$. The typical shape of 
the distribution can be obtained by rescaling this probability by $n$.} 
Since we are interested in the logarithmic shape of the distribution, we 
disregard any multiplicative constant. As $n\to \infty$, \john{we thus 
obtain that the probability } profile converges towards \john{a} universal 
limit \john{independent of the particular shape of the distribution $\rho$, 
precisely given by:}
\[p_{\tau}^{0,\infty} \propto \int_{0}^1 x(1-x)^\tau dx = \frac{1}{(\tau+1)(\tau+2)}\]
which is indeed a power law with exponent $-2$, identical to the one arising in
critical systems, consistent with those reported in
neuronal data~\cite{friedman2012universal}, in neuron models (Fig.~\ref{fig:Brunel})
and in surrogate systems (Fig.~\ref{fig:Poisson}).

\begin{figure}[htbp]
\centering
\includegraphics[width=.3\textwidth]{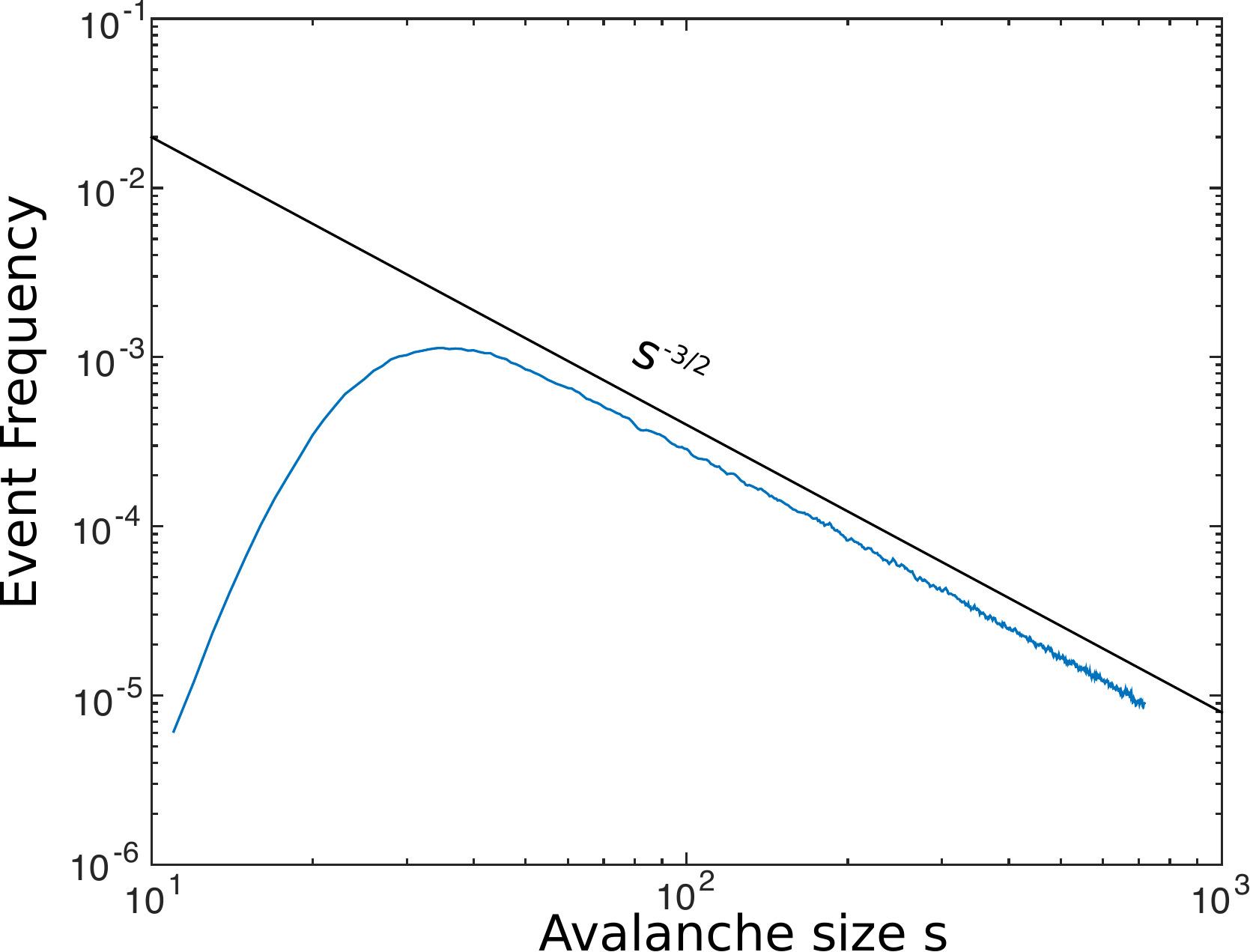}
\caption{Universal shape of the avalanche size distribution in the slow rate and large $n$ limit.}
\label{fig:UniversalShape}
\end{figure}

We now show that this extends to the distribution of avalanche size and the scaling of the mean avalanche size with duration.
Indeed, the size of avalanches of duration $\tau$ have a binomial
distribution, corresponding to $s-\tau$ successes among
$(n-1)\tau$ independent Bernoulli variables with
probability of success $1-q(t)$. Moreover, De Moivre-Laplace
theorem~\cite{feller2008introduction} ensures convergence as $n$
increases towards a normal variable with mean
$(n-1)\tau p(t)$ and variance $(n-1)\tau p(t)q(t)$ with $p(t)=1-q(t)$.
Using again our separation of timescale and stationarity hypotheses, we
find the probability the probability of finding an
avalanche of size $s$ and duration $\tau$ averaged on the firing rate:
\[
	\bar{p}^n(s,\tau)\sim \int_{0}^1 \;e^{-\frac{(s-\tau-n (1-q) \tau)^2}{2n q \tau(1-q)}} \frac{q^{2n}(1-q^n)^\tau}{\sqrt{2\pi n q \tau(1-q)}} \rho(q)dq\]
  which converges, as $n\to \infty$, towards:
  \[
  	\bar{p}^{\infty}(s,\tau)\sim   \int_{0}^1 e^{\frac{(s-\tau(1+\log(u)))^2}{2\tau\log(u)}}\frac{u(1-u)^\tau}{\sqrt{2\pi\tau\log(u)}}\; du
\]
We thus obtain at leading order the size distribution:
\begin{align}
\mathcal{P}(s) 
&\sim e^s \sum_{\tau=1}^s \int_{0}^1 e^{\frac{(s-\tau)^2}{2\tau\log(u)}}\frac{u(\sqrt{u}(1-u))^\tau}{\sqrt{2\pi\tau\log(u)}}\; dx.
\end{align}
It is hard to further simplify this formula, but it can be easily evaluated numerically. We depict the result of this computation in Fig.~\ref{fig:UniversalShape}, and illustrate the apparent power-law scaling with slope $-3/2$.

Eventually, we obtain for the average size $A_{\tau}$ of avalanches of duration $\tau$:
\begin{align*}
A_{\tau} &\sim \int_{0}^1 \int_{s=\tau}^{\infty} s \; e^{\frac{(s-\tau(1+\log(u)))^2}{2\tau\log(u)}}\frac{ds}{\sqrt{2\pi\tau\log(u)}}\; du\\
&\sim \tau^{3/2} \int_{0}^1 (-\log(u))^{5/4} u^{\sqrt{\frac{\tau}{-8\pi\log(u)}}}\;du.
\end{align*}
We thus conclude that while a power-law relationship persists between $A_{\tau}$ and
$\tau$, the scaling exponent is not related to the exponents of the power-law
of size ($3/2$) and duration ($2$) distribution through Sethna's crackling noise
relationship, which would predict an exponent equal to $2$. Importantly, we note
that the exponent found here is quantitatively perfectly consistent with the
exponent found in \emph{in vitro} data~\cite{friedman2012universal} in the neural
network model (Fig.~\ref{fig:Brunel}) or surrogate Poisson system (Fig.~\ref{fig:Poisson}).

\section{{Spike pattern entropy and information capacity}}
{We have thus proved that power-law distributions of avalanches do not necessarily
reveal that the network is operating at criticality. However, a number of theories 
have proposed that operating at criticality was an optimal regime of information processing
in the brain, maybe selected by evolution as a useful trait for the nervous system
~\cite{shew2011information,hesse2015self,shew2009neuronal,shew2011information,shew2013functional,
beggs2012being,beggs:08}. The question that arises is thus whether these theories break down
when power-law statistics no more arise from the system operating at criticality, but from 
a mean-field Boltzmann chaos regime.}

{In order to address this outstanding question, we came back to the methods used in order 
to demonstrate optimality of data processing capabilities at criticality. These theories 
rely on the computation of the information capacity of the network in different activity 
regimes evaluated, following Shannon's information theory,
as the entropy of the patterns of spike fired. In detail, a spike pattern in a network of size
$N$ is a $N$-uplet $s\in \{0,1\}^{N}$, with $s_{i}=1$ (resp. $s_{i}=0$) if neuron $i$ has fired (resp., not fired) 
in a specific timebin. If $p$ denotes the probability of occurrence of spike patterns, the entropy is given by:
\[\textrm{Entropy}=\sum_{s\in\{0,1\}^{N}} p(s) \log(p(s)).\]
}

\begin{figure}
\includegraphics[width=.45\textwidth]{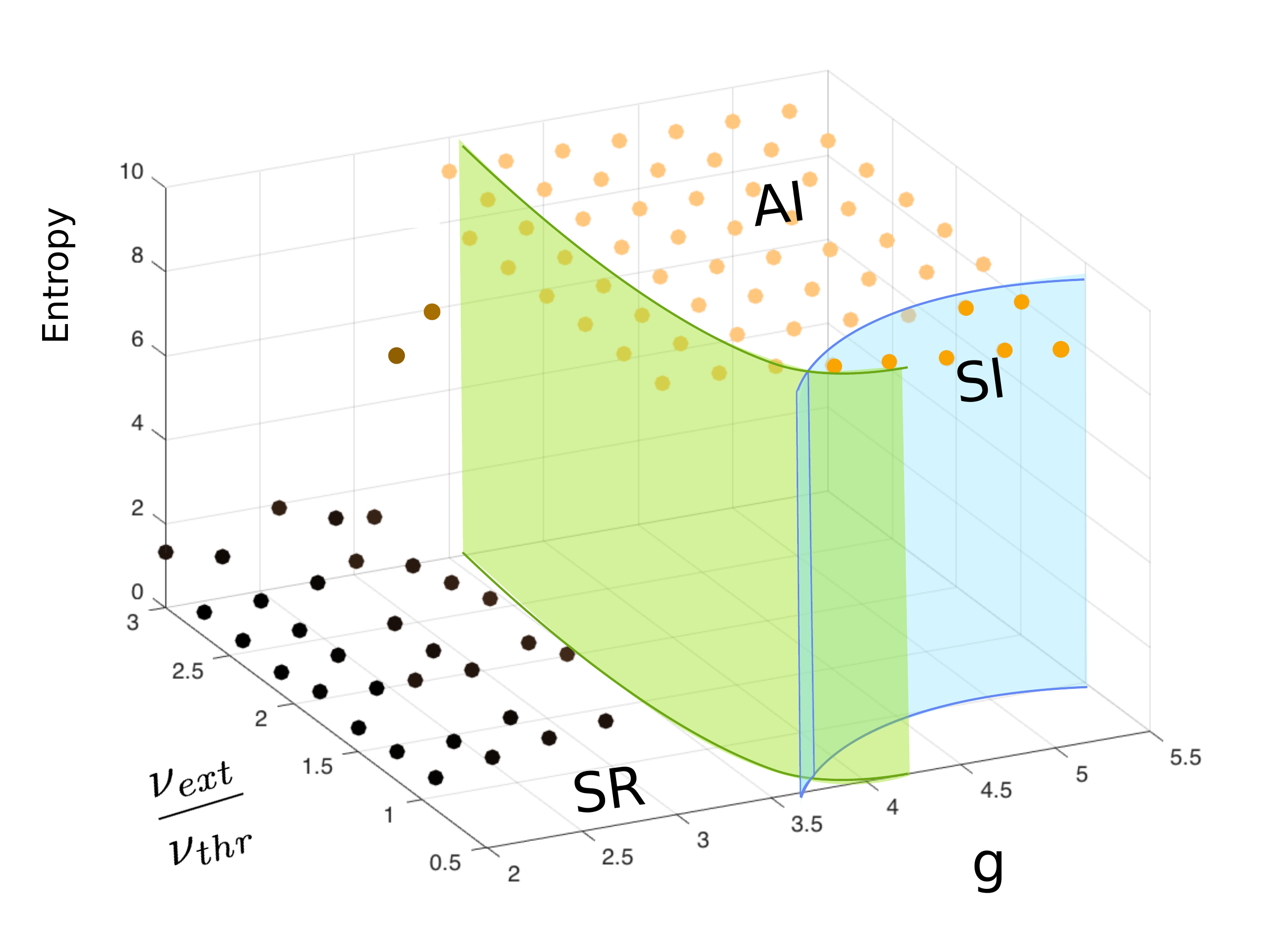}
\caption{{Computation of the entropy (circles) of the network in the Brunel's model~\cite{brunel2000dynamics} for 
distinct values of the input intensity $\nu_{ext}$ and inhibition ratio $g$. Color code indicates the entropy 
amplitude. Parameters as in the original model~\cite[Fig.~2B]{brunel2000dynamics}. The blue surface represents 
the boundary between SI and AI regimes, and the green surface separates
SR regimes from SI regimes. We observe a clear transition from low entropy (SR) to high entropy (SI+AI). }}
\label{fig:Entropy}
\end{figure}

{In order to test whether our theory accounting 
for the emergence of power-law distributions in the absence of criticality 
challenges high information capacity of neuronal networks,
we computed the information capacity of Brunel's model in different regimes (see Fig.~\ref{fig:Entropy}). 
The numerical results show that this is not the case, and the information capacity 
is maximal in the SI regime where power-law statistics of avalanches were observed. 
However, we observe no difference between entropy levels in the SI or AI states. Therefore,
we conclude that the maximality of entropy is not necessarily related to the emergence of 
power-law statistics.}

{These observations can be well understood heuristically. Indeed, the entropy
of spike patterns is a measure of the variability of possible spike patterns 
observed in the course of neuronal activity. The diagram of Fig.~\ref{fig:Entropy}
is thus not surprising. Indeed, while the diversity of spike patterns is reduced in the 
highly synchronized regular regimes, it will be large in the irregular regimes, both synchronized 
asynchronous~\footnote{In the 
models based on branching processes with parameter $p$, it is clear that the maximal diversity
arises at criticality, where the entropy is larger than in the sub-critical regime (only patterns
with a small number of spikes) or in the super-critical regime $p>1$ (patterns with a large number 
of spikes).}. In other words, entropy is maximized within irregular regimes where 
more diverse patterns are fired, and this independently of the underlying 
mechanisms supporting the emergence of the irregular activity.}

 \section{Local field potentials are ambiguous measures of criticality}\label{sec:LFP}

 We have thus shown that discarding the criticality assumption does not 
 degrade the quantifications of network efficiency This being said, we are facing an apparent contradiction. Indeed, our
 theory provides an account for the presence of critical statistics in
 networks in the SI regime, but discards AI states as possibly having
 critically distributed avalanches. Evidences of critical statistics of
 avalanches \emph{in-vivo} in the awake brain have been {reported}, but
 are more scarce and controversial.  Unlike neuronal cultures, the
 activity in the awake brain does not display bursts separated by
 silences, but is sustained. Using a macroscopic measurement of
 neuronal activity, the Local Field Potential (LFP), power laws could
 be shown from the distribution of peaked
 events~\cite{petermann2009spontaneous}.  The motivation to use
   negative LFP peaks to deduce information from the distribution
   of spike avalanches relied on the fact that the amplitude of these 
   peaks correlated with firing activity~\cite{shew2013functional,beggs2003neuronal}.

However, this monotonic relationship between the number of spikes and the 
number of spikes do not imply that there should exist a relationship 
between the distribution of peak amplitude and avalanches. Moreover, 
it was shown that power-laws naturally emerge from
 the random nature of the signal and the thresholding procedure used in
 this analysis, and moreover these power-laws may not be statistically
 significant~\cite{touboul2010can}. Indeed, no power-law
   scaling could be found from unit activity, which were better fit by
   double-exponential distributions~\cite{dehghani2012avalanche}.
   These analyses rather suggest that the power-law statistics of LFP
   peaks does not reflect scale-invariant neural activity.

 In an attempt to clarify this, we investigate here whether the
   distribution of LFP peaks can display power-law scaling in spiking
   networks or in their stochastic surrogates. To obtain a more
   biophysical model where LFP can be defined, we considered the
 current-based Vogels and Abbott model~\cite{vogels2005signal} which
 provides a biologically realistic model of spiking network
 displaying asynchronous irregular and synchronous regular
   states, and in which synaptic currents are described by exponentially
   decaying functions with excitation and inhibition having distinct time 
   constants (in place of Dirac impulses in the Brunel model~\eqref{eq:Brunel} 
   see~\cite{vogels2005signal} for details). Simulations of the model 
   provide an instantaneous distribution of postsynaptic currents, from 
   which we computed LFP signals. 
   
\begin{figure}[htbp]
 	\centering
 		\includegraphics[width=.5\textwidth]{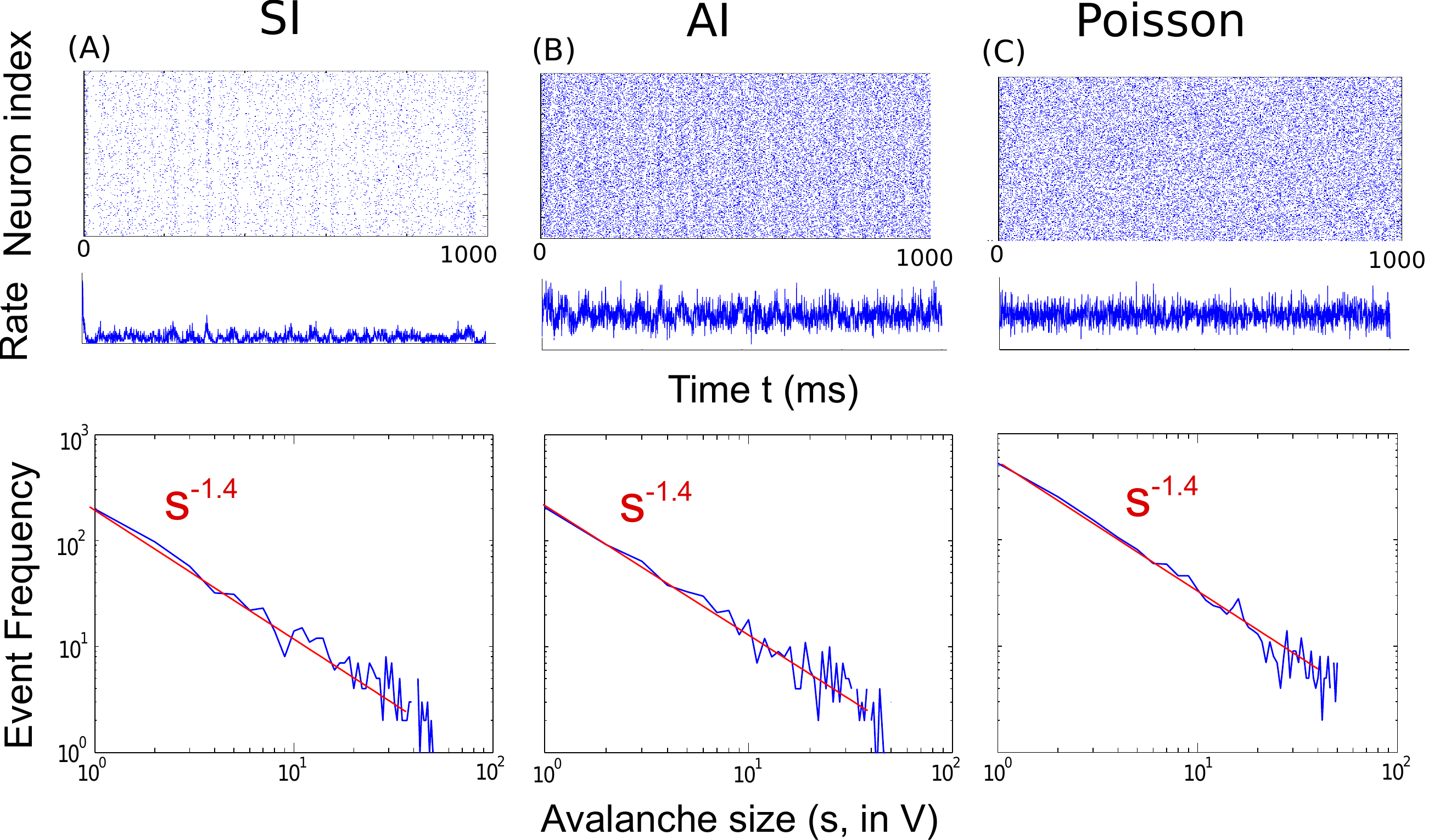}
             \caption{Avalanche analysis defined from the macroscopic
               variable $V_{LFP}$ of a network of integrate-and-fire
               neurons with exponential synapses. Networks displaying
               an SI (left) or AI (middle) state, and a purely 
               stochastic surrogate (right), show similar macroscopic
               power-law LFP peak statistics with the same exponent close to -3/2.}
 	\label{fig:VA}
 \end{figure}

   In detail, we have considered a spatially extended neural network of $5\,000$
   units randomly located on a 2-dimensional square and satisfying the 
   by the Vogels-Abbott model. We evaluated an LFP signal $V_{LFP}$
   from the postsynaptic currents according to Coulomb's law~\cite{nunez2006electric}:
 $$ V_{LFP} = {R_e \over 4 \pi} \ \sum_{j} { I_j \over r_j } ~ ,$$
 where $V_{LFP}$ is the electric potential at the electrode position,
 $R_e$ = 230~$\Omega$cm is the extracellular resistivity of brain
 tissue, $I_j$ are the synaptic currents of and $r_j$ is the distance
 between $I_j$ and electrode position. Remarkably, applying to this
 more sophisticate model the same procedures as in the original
 paper~\cite{petermann2009spontaneous}, we found that the method cannot
 distinguish between structured or non-structured activity: for a fixed
 firing rate and bin size, we have been comparing in
 Figure~\ref{fig:VA} the LFP statistics of the avalanche duration in a
 network of neurons in the AI regime (A) or in the SI regime (C) that
 shows partial order of the firing and saw no difference. We also
 compared the statistics to those of independent Poisson processes with
 constant rate (B): the three instances show power-law scaling of
 avalanche duration, with the same exponent, that seem to rather be
 related to the firing rate and bin size than to the form of the
 network activity.  The scaling coefficient is again, close from $3/2$,
 and varies with bin size. Clearly, in this case, the dynamics
   of LFP peaks cannot distinguish between critical and non-critical
   regimes.

{\section{Discussion}}

{In this paper, we have evaluated how power-law statistics and
  universal scaling can arise in the absence of criticality.  We first
  outline the novel contributions of the present manuscript, then we
discuss their significance.}

{The contributions of the present manuscript are the following: 
\renewcommand{\theenumi}{(\roman{enumi})}
\begin{enumerate}
\item We first investigated the avalanche statistics of spiking neural
  networks models (Section~\ref{sec:aval}).  We showed that these
  networks display power-law statistics with the critical exponents
  -3/2 and -2, as well as collapse of the shapes of avalanches.  This
  observation is robust and valid for a wide region of parameters
  corresponding to synchronous irregular activity in such networks,
  and thus away from any transition point (i.e., away from
  criticality).
\item This robust numerical observation led us to investigate
  theoretically how such statistics can emerge from the collective
  activity of spiking networks. One ubiquitous property of large
  neural networks models is that neurons behave as independent
  processes with the same statistics. This property, termed
  \emph{Boltzmann's molecular chaos} regime in statistical physics, or
  \emph{propagation of chaos} in mathematics, is universal in the
  dynamics of large-scale networks. We reviewed such properties in
  models, as well as recent experimental evidence supporting this
  decorrelation between the dynamics of cells.
\item We next tested the hypothesis that power-laws may emerge from
  such large systems of weakly correlated units with similar
  statistics. We introduced and investigated numerically the dynamics
  of a surrogate network made of independent neurons sharing the same
  statistics (Section~\ref{sec:surrog}). Surprisingly, despite the
  simplicity of these systems, they display precisely the same
  power-law statistics of avalanche duration and size with the same
  exponents and shape collapse as fully-connected networks.
\item This surrogate model is simple enough to derive in closed-form
  the statistics of the duration and size of the avalanches.  Using
  this analytic expression, we demonstrate that the `critical'
  exponents emerge naturally in large-scale systems (operating 
  within the Boltzmann chaos regime), without the
  need to invoke criticality (Section~\ref{sec:analytics}).
\item Current literature indicate that criticality is an optimal 
  information processing regime for the brain. We considered the 
  same measures of information capability (entropy of spike trains) 
  on our biologically 
  plausible models and showed that indeed, information is maximized 
  in the SI regime where power-laws emerge. However, we show that
  this is not exclusive of the SI regime but same levels of entropy
  are found in AI regimes where no avalanche can be defined. 
\item Finally, we also addressed the presence of power-law scalings
  observed in LFP recordings by simulating such signals emerging from
  more realistic neuronal network models with excitatory and
  inhibitory cells. Surprisingly, we observed that power-laws with the
  same exponents as observed in experiments are found in all regimes
  tested.  This exponents persisted when neurons were replaced by
  independent Poisson processes. This clearly indicates that power-law
  scalings in LFPs do not constitute any proof of criticality in the
  underlying system.
\end{enumerate}}

{All together, these numerical and theoretical findings provide a
  new interpretation for the emergence of power-law statistics in
  large-scale systems, independent of the notion of criticality.
  {We propose to explain the emergence of such scaling based on
    Boltzmann molecular chaos regime, known to govern the dynamics of
    most large-scale interacting systems~\cite{sznitman:89,brunel2000dynamics,touboul2014propagation}.
    In other words, power-law and universal scaling functions can be
    due to a mean-field effect in systems made of a large number of
    interacting units.  Of course, this theory does not hypothesize
    that the elements considered (here, neurons) are disconnected in
    reality. To the contrary, the fact that the critical exponents
    still resist the removal of interconnections shows that such
    exponents do not need criticality to be explained.}}

{The main mechanism explored here, Boltzmann's molecular chaos,
  is a universal feature of many statistical systems.} The very
particular structure of different particles activity it induces,
namely statistical independence of the particles behavior together
with a correlation in the law, may induce as we have observed, the
same type of power-laws as in critical systems, with universal
coefficients that are consistent with {those found in} critical
systems. In agreement with this theory, we have seen that the same
statistics are reproduced by a sparsely connected network and a
surrogate stochastic process where the periods of firing and silences
are themselves generated by another stochastic process. This interpretation 
suggests that similar scaling relationships shall
arise in more realistic neural network models with fixed connectivity 
patterns, in particular including 
axonal propagation delays (constant delays are already present in Brunel
model), dendritic structure, spike frequency adaptation,
non-instantaneous synaptic transmission. Indeed, most of these elements
will make the intrinsic dynamics of each cell more complex, but we do not  
expect that this complexity could affect the fact that these systems operate
within the Boltzmann molecular chaos regime. Notwithstanding, models including
synaptic plasticity, which is the process by which the brain acquires skills and 
stores memories, may not belong to the class of systems described in this paper. 
Indeed, in such systems, the connectivity patterns vary depending on the 
pairwise correlations of cells activity, and this relationship may compete 
with the establishment of Boltzmann's molecular chaos regime. While this may 
not occur in the adult brain where plasticity is much slower than 
neuronal activity, distinct phenomena not described by our model may occur
occur during the critical periods of brain development, when plasticity occurs
at a faster timescale. Further experimental 
and theoretical investigations are necessary to characterize avalanche distributions
in these systems, as well as to compute correlation levels to test if the decorrelation 
characteristic of Boltzmann's molecular chaos occurs.

Interestingly, a network operating in Boltzmann's molecular chaos regime can be
interpreted as a high dimensional system with hidden variables, as
studied recently in~\cite{aitchison2014zipf,schwab2014zipf}. In these
contributions, the authors investigate the rank distribution of high
dimensional data with hidden latent variable, and show that such
systems display Zipf law scaling (power-laws with slope $-1$ in the
rank distribution) that generically arise from entropy consideration,
and using the elegant identity between entropy and energy shown
in~\cite{mora-bialek:11}.  While these developments do not generalize
here, Boltzmann molecular chaos provides a natural explanation for the
emergence of weakly correlated units with similar probability laws: in
the neural network system case, the common rate could be seen as a
latent variable, and both independence and irregularity build up only
from the interactions between cells. As a result of this theory,
revealing apparent power-law scaling with exponents of -3/2 and 2, as
well as shape collapse, may be entirely explained statistically; in
particular, these criteria constitute no proof of criticality and
experimental studies solely relying on them should be re-evaluated.

This statement is even more true when it comes to macroscopic
{measurements such as the LFP}: both SI and AI regimes, as well
as stochastic surrogates, display power-law statistics in the
distribution of LFP peaks. This shows that systems of weakly
correlated units, or their stochastic surrogates, can generate
power-law statistics when considered macroscopically.  {Here
  again,} the power-law statistics tells nothing about the critical or
non-critical nature of the underlying system. This potentially
reconciles contradictory observations that macroscopic brain variables
display power-law scaling~\cite{petermann2009spontaneous}, while no
sign of such power-law scaling was found in the
units~\cite{dehghani2012avalanche,bedard2006does}.  More generally,
these results also put caution on the interpretation of power-law
relations found in nature.

{A question that naturally emerges {is} how to distinguish
  power-laws {due to} criticality from those {due to
    Boltzmann's} molecular chaos regime.} {We used the previous
  observation~\cite{sethna2001crackling} that a prominent
  characteristic of criticality beyond the presence of power-law
  scalings is the particular relationship one finds between the
  exponents.  We found here that the power-law scalings emerging in
  the absence of criticality did not satisfy this relationship.  We
  propose to use that criterion as a possible way to distinguish
  between power-law scaling due to criticality or due to Boltzmann's
  molecular chaos.}

{In conclusion, we have shown here that stochastic models can
  replicate many of the experimental observations about 'critical'
  exponents, which demonstrates that not only power-law scaling is not
  enough to prove criticality, but that we need new and better methods
  to investigate this in experimental systems.  The fact that that
  such exponents are seen for networks and for stochastic systems,
  shows that they apply to a large class of natural systems and may be
  more universal than previously thought.}  As Georges Miller noticed
in his seminal paper~\cite{miller1957some}, examining random text
typed by virtual monkeys, the texts produced may not be interesting,
but have \emph{some of the statistical properties considered
  interesting when humans, rather than monkeys, hit the keys.}
{Similarly, the present results show systems that can emulate
  the power-law scaling seen in brain activity, but with no
  criticality involved.  We thus cannot conclude on whether the brain
  operates at criticality or not, but we need more elaborate methods
  to resolve this point.}

\subsection*{Acknowledgments}

A.D. was supported by the CNRS and grants from the European Community
({\it BrainScales} FP7-269921 and {\it Human Brain Project}
FP7-604102).  We warmly thank Quan Shi and Roberto Zu\~niga Valladares
for preliminary work and analyzes. \john{We thank anonymous referees
  for their suggestions of analyses and references.}

\appendix
\section{Power law statistics and maximum likelihood fits for stationary data}
We review here the methods used to fit the power law distributions, that closely follow the methodology exposed in~\cite{clauset2009power}
This methodology applies to stationary data. Taking the logarithm of the probability density of a power-law random variable,
we obtain $\log(p(x)) = -\alpha \log(x) + \log(a)$. The histogram of the
power-law therefore presents an affine relation in a log-log plot. For this reason, power-laws in empirical data are often studied
by plotting the logarithm of
the histogram as a function of the logarithm of the values of the random
variable, and doing a linear regression to fit an affine line to through
the data points (usually using a least-squares algorithm).  This method
dates back to Pareto in the 19th century (see e.g.  \cite{arnold:83b}).  The
evaluated point $\hat{x}_{\min}$ corresponding to the point where the data start
having a power-law distribution is mostly evaluated visually, but this method is
very sensitive to noise (see e.g.
\cite{stoev:06} and references herein). The maximum likelihood estimator of the exponent
parameter $\alpha$ corresponding to $n$ data points $x_{i}\geq {x_{\min}}$
is:
\[\hat{\alpha} = 1+n \Big( \sum_{i=1}^{n}
\log \frac{x_{i}}{{x_{\min}}}\Big)^{{-1}}.\]
The log-likelihood of the data for the estimated parameter value is:
\[L(\hat{\alpha} \vert
X)  = n\, \log \left(\frac{\hat{\alpha}-1}{{x_{\min}}}\right) - \hat{\alpha}
\sum_{i=1}^{n}\log\left ( \frac{x_{i}}{{x_{\min}}} \right).\]

The parameter $\hat{x}_{\min}$ is evaluated then by
minimizing the Kolmogorov--Smirnov distance:
\[KS=\max_{x\geq {x_{\min}}} \vert S(x)-\hat{P}(x) \vert \]
where $S(x)$ is the cumulative distribution function (CDF) of
the data and $\hat{P}(x)$ is the CDF of the theoretical distribution
being fitted for the parameter that best fits the data for $x\geq
{x_{\min}}$), as proposed by Clauset and colleagues in
\cite{clauset-etal:07}. In order to quantify the accuracy of the fit, we use a standard goodness-of-fit test
which generates a p-value. This quantity characterizes the likelihood of obtaining a fit as
good or better than that observed, if the hypothesized distribution is correct.
This method involves sampling the fitted distribution to generate
artificial data sets of size $n$, and then calculating the Kolmogorov--Smirnov
distance between each data-set and the fitted distribution, producing the
distribution of Kolmogorov--Smirnov distances expected if the fitted distribution
is the true distribution of the data.  A p-value is then calculated as the
proportion of artificial data showing a poorer fit than fitting the observed data
set. When this value is close to $1$, the data set can be considered to be drawn
from the fitted distribution, and if not, the hypothesis might be
rejected. The smallest p-values often considered to validate the statistical test
are taken between $0.1$ and $0.01$.  These values are computed following
the method described in \cite{clauset-etal:09}, which in particular
involves generating artificial samples through a Monte-Carlo procedure.

These methods, very efficient for stationary data, fail to evaluate the tails of non-stationary data
as is the case of neuronal data. A weighted Kolmogorov-Smirnov test with a refined goodness of fit estimate
valid up to extreme tails~\cite{chicheportiche2012weighted}.

\section{Subsampling effects}
Of course, any analysis of finite sequences of data is subject to subsampling effects. While these may be neglected for light-tailed data, they become prominent when it comes to assessing possible slow decay of the tails of  a statistical sample distribution. These effects were discussed in detail in a number of contributions. In the context of neuronal avalanches, these effects were characterized in~\cite{priesemann2009subsampling}, and the results show indeed a modification of the slope with subsampling, together with exponential cutoffs pushed to larger sizes as sampling becomes finer.
\begin{figure}[!h]
	\centering
		\includegraphics[width=.5\textwidth]{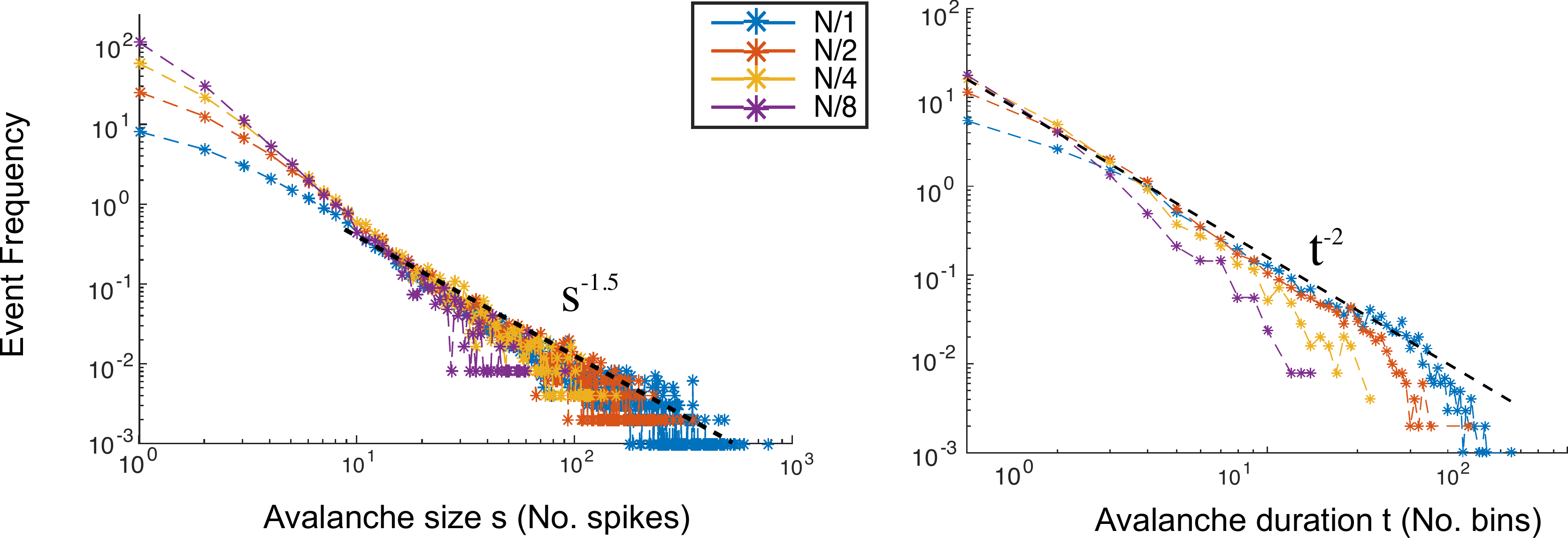}
	\caption{Subsampling effects. Statistics for a randomly extracted subset of neurons of size $n=125, 250, 500$ among $1\,000$ neurons whose dynamics is described by Brunel's model. Theoretical power-laws with critical exponents are displayed in black dashed lines.}
	\label{fig:subsampling}
\end{figure}

We have confirmed these results in our own data. In Fig.~\ref{fig:subsampling}, we have computed the distribution of avalanche size and duration when considering only a fraction of the neurons for the statistics. In detail, we have simulated the Brunel model~\cite{brunel2000dynamics} with $N=1\,000$. This yields a raster plot, from which we have extracted a randomly chosen subset of $n$ neurons, with $n=N/k$ for $k\in \{2,4,8\}$. We indeed observed that an exponential cutoff is shifted towards larger sizes and slopes increase with the subsampling ratio $k$.

\section{Brunel's model}
In our simulations, we have used the neuronal network model introduced by Brunel in~\cite{brunel2000dynamics} and have referred to the different dynamical regimes of this system. We review here the model, provide all parameters used in our simulations and show that the conclusions drawn in one example of the synchronous irregular (SI) state are valid for all parameters tested within this regime.
\begin{figure}[!h]
	\centering
		\includegraphics[width=.5\textwidth]{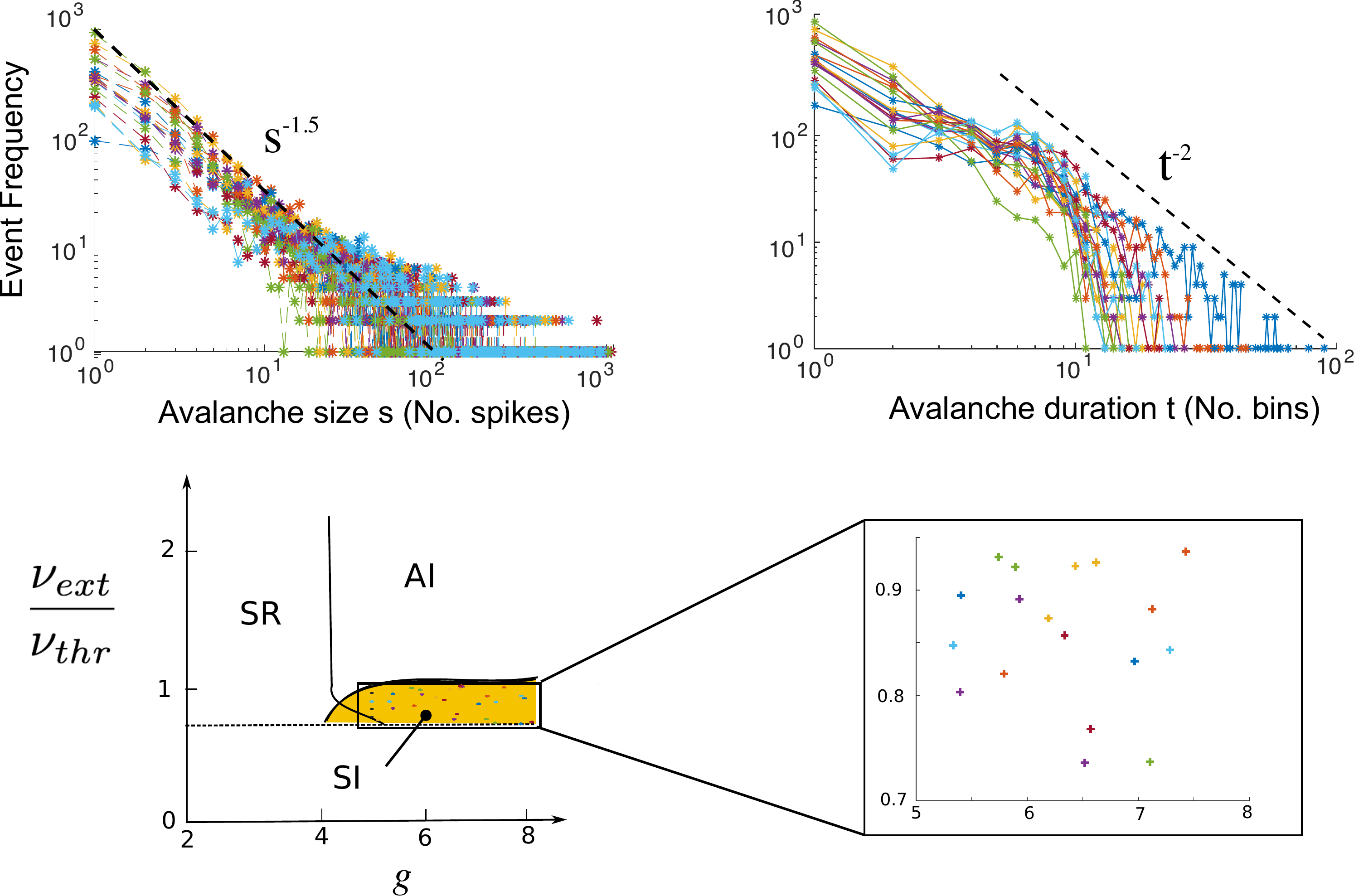}
	\caption{Avalanche statistics for the Brunel model with randomly chosen parameters within the SI regime. }
	\label{fig:BrunelAllSIs}
\end{figure}
The model describes the dynamics of $N$ integrate-and-fire neurons, $80\%$ of which are excitatory and the others inhibitory. In the model, it is assumed that each neuron receives $C=\varepsilon N$ randomly chosen connections, that are assumed to uniformly arise from the excitatory and the inhibitory population, thus  $80\%$ of the incoming connections to any cell come from the excitatory population. The network is assumed to be sparsely connected, thus $\varepsilon\ll 1$. The depolarization $v^i$ of neuron $i$ at the soma satisfies the equation:
\[\tau \der{v^i}{t}=-v^i +RI_i(t)\]
where $I_i(t)$ is the total current reaching the soma at time $t$. These currents arrive from the synapses made with other cells within the network, as well as from connections to neurons outside the network. It is assumed that each neuron receives $C_{ext}$ connections to from excitatory neurons outside the network, and that these synapses are activated by independent Poisson processes with rate $\nu_{ext}$. The current receives by neuron $i$ is thus the sum:
\[RI_i(t)=\tau \sum_{j=1}^{C+C_{ext}} J_{ij}\sum_{k}\delta(t-t_j^k-D)\]
where the sum is taken over all synapses, $J_{ij}$ are the synaptic efficacies, $t_j^k$ are the spike times at synapse $j$ of neuron $i$, and $D$ is the typical transmission delay, considered homogeneous at all synapses for simplicity. In order to simplify further the model, it is assumed that $J_{ij}=J>0$ for all excitatory synapses, and $J_{ij}=-gJ<0$ for inhibitory synapses. The parameter $g$ is relevant in that it controls the balance between excitation and inhibition: if $g<4$, the network is dominated by excitation, and otherwise it is dominated by inhibition.
The neuron $i$ fires an action potential when $v^i$ reaches a fixed threshold $\theta$, and the depolarization of neuron $i$ is instantaneously reset to a fixed value $V_r$ where it remains fixed during a refractory period $\tau_{rp}$ (during this period, the neuron is insensitive to any stimulation). An important parameter is the ratio between the rate of external input $\nu_{ext}$ and the quantity denoted $\nu_{thresh}$ corresponding to the minimal frequency that can drive one neuron, disconnected from the network, to fire an action potential: $\nu_{thresh}=\frac{\theta}{0.8 J \tau}$ (the coefficient $0.8$ in that formula corresponds to the fraction of excitatory neurons).

In this model, the parameters that are kept free are the balance between excitation and inhibition $g$ and the external firing rate $\nu_{ext}$. All other parameters are chosen as in table~\ref{tab:paramBrunel}
\begin{table}
	\begin{center}
	\begin{tabular}{|c|c|c|c|c|c|}
		\hline
		$\varepsilon$ & $D$ & $J$ & $\tau_{rp}$ & $\theta$ & $V_r$\\
		\hline
		 0.1 & 1.8 ms & 0.2 mV & 2 ms & 20 mV & 10 mV\\
		\hline
	\end{tabular}
	\end{center}
	\caption{Parameters used in all simulations of Brunel's model, as in~\cite{brunel2000dynamics}.}
	\label{tab:paramBrunel}
\end{table}
Using a mean-field analysis together with a diffusion approximation, the authors find that all neurons are independent point process driven by a common rate $\nu(t)$ given by a self-consistent equation. Heuristically, during the time interval $[t,t+dt]$, the probability for any given to spike is given by $\nu(t)\,dt$, and the realization of this random variable are independent in the different neurons. When the rate $\nu(t)$ depends on time, neurons thus show a level of synchrony, and when $\nu(t)$ is constant, the regime is called asynchronous, in the parlance of~\cite{brunel2000dynamics}. In that paper, an analysis of the self-consistent rate equation in the mean field limit led to the identification of several regimes that are depicted in Fig.~\ref{fig:BrunelAllSIs}:
\begin{itemize}
	\item The asynchronous irregular (AI) state in which $\nu(t)$ converges towards a strictly positive constant value, which occurs when the excitatory external inputs are sufficiently large ($\nu_{ext}>\nu_{thresh}$) and when inhibition dominates excitation.
	\item The synchronous regular (SR) regime corresponds to a state in which $\nu(t)$ is a periodic function of time. This regimes arises in the excitation dominated regime, and the oscillations frequency is controlled essentially by the transmission delay $D$ and the refractory period $\tau_{rp}$ (approximately varying as $\tau_{rp}/D$). The transition thus occur close from the line $g=4$.
	\item The synchronous irregular (SI) regime occurs essentially in the inhibition-dominated regime when the input are not sufficient to drive the network to a sustained firing state, i.e. when $\nu_{ext}<\nu_{thresh}$.
\end{itemize}
We have reproduced in Fig.~\ref{fig:BrunelAllSIs} the bifurcation diagram~\cite[Fig.2 B]{brunel2000dynamics} with the bifurcation lines between AI, SI and SR states. Within the SI state, we have been randomly drawing 30 parameter points and analyzed the avalanches arising for these parameters. We have found that all regimes show a very clear power-law distribution of avalanche size and duration with exponents consistent with the exponents $-1.5$ and $-2$ predicted by the theory.

\section{No slowing down within the SI regime}

\john{In addition to the fact that the SI regime is away from any
  transition between the different network regimes, we confirmed that
  the system did not show the typical properties of critical states. A
  number of criteria were proposed in the Ising model to be
  characteristic of the critical regime. These include the divergence
  of the correlation length, of the heat capacity or magnetic
  susceptibility, that are all related to long-range correlations
  between spins. Here, the absence of order parameter and spatial
  dimension prevents from using similar criteria to investigate the
  presence of critical dynamics. However, a criterion independent of
  the definition of an analogous of order parameter or magnetic
  susceptibility is the \emph{critical slowing-down} occurring at
  phase transitions for dynamical systems. This criterion states that
  the relaxation time of the system, namely the time it takes for the
  system to return to its stationary regime after a perturbation,
  diverges at criticality.}

\begin{figure*}
\includegraphics[width=.8\textwidth]{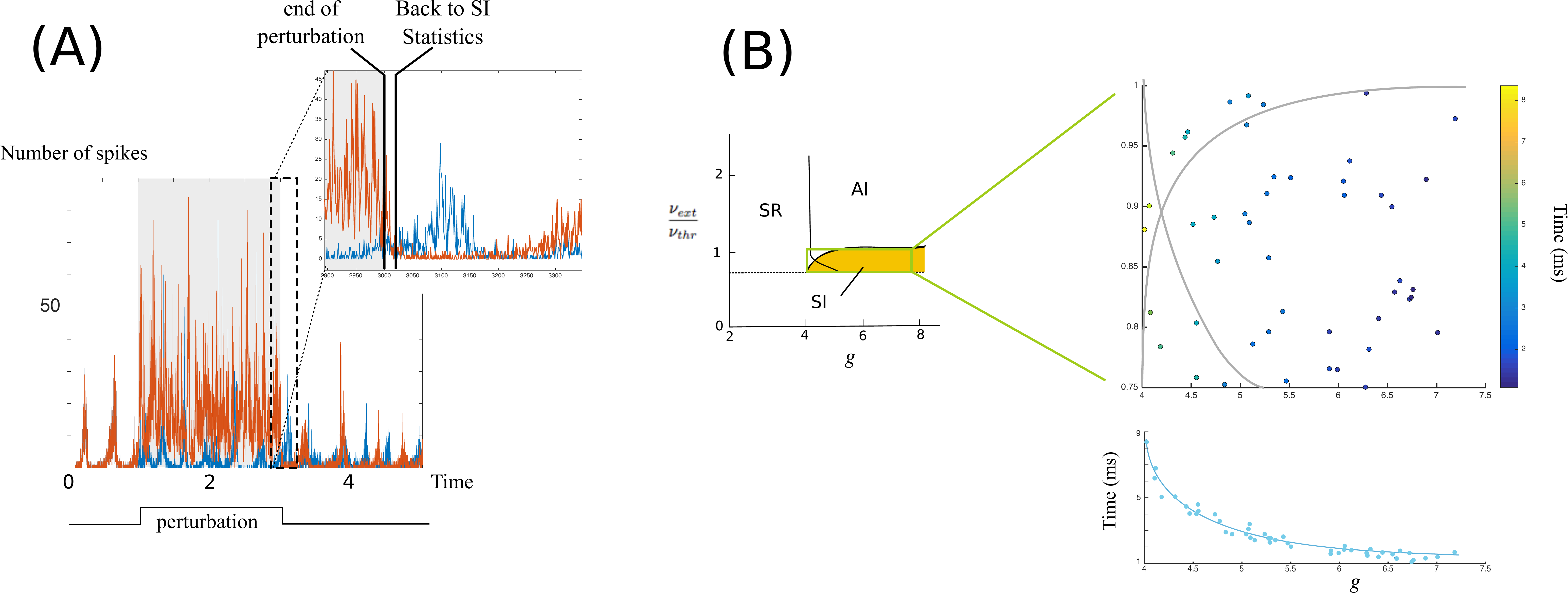}
\caption{\john{Relaxation time of the Brunel model in the SI regime. 
(A) Typical trajectory (blue) perturbed by a constant input (red) into 
a SI regime, returns to the SI state after a few milliseconds. 
This is true over the whole SI domain (B): relaxation times are on 
the order of a few milliseconds, and increase sharply close from the 
transition.}}
\label{fig:SlowDown}
\end{figure*}

\john{In the present case, computing relaxation times is a challenge
  since the system is not at an equilibrium but within a chaotic
  regime, thus all perturbations produce massive changes in the
  dynamics of the system. Following the methodology developed in~\cite{
  hansel:16,zerlaut}, we designed a numerical criterion to evaluate
  relaxation time to the SI regime. That regime is essentially defined by the
  alternation of periods of collective activity followed by silences.
  We have thus perturbed the system by adding a constant input within
  a short time window (See Fig.~\ref{fig:SlowDown}(A)), which has the
  effect of switching the system into an asynchronous irregular regime
  where the firing is uninterrupted.  As the perturbation stops, the
  system quickly returns to an SI regime with alternations of silences
  and collective bursts.  An upper bound of the relaxation time can
  thus be defined as the first time, after the perturbation has
  stopped, at which the system is completely silent. We have made
  extensive simulations within the SI regime to compute the relaxation
  time and obtained that the system returns to SI statistics after a
  few milliseconds after stimulation (on the order of 2ms). This time
  increases very fast close to the SI-AI transition as expected from
  the theory, but within the SI regime, the system did not show any
  indication of critical slowing-down.  }

\

\section{Diverse regimes of independent processes}
We have confirmed that the statistics of independent Poisson processes with fluctuating instantaneous firing
\begin{figure}[h]
	\centering
		\includegraphics[width=.4\textwidth]{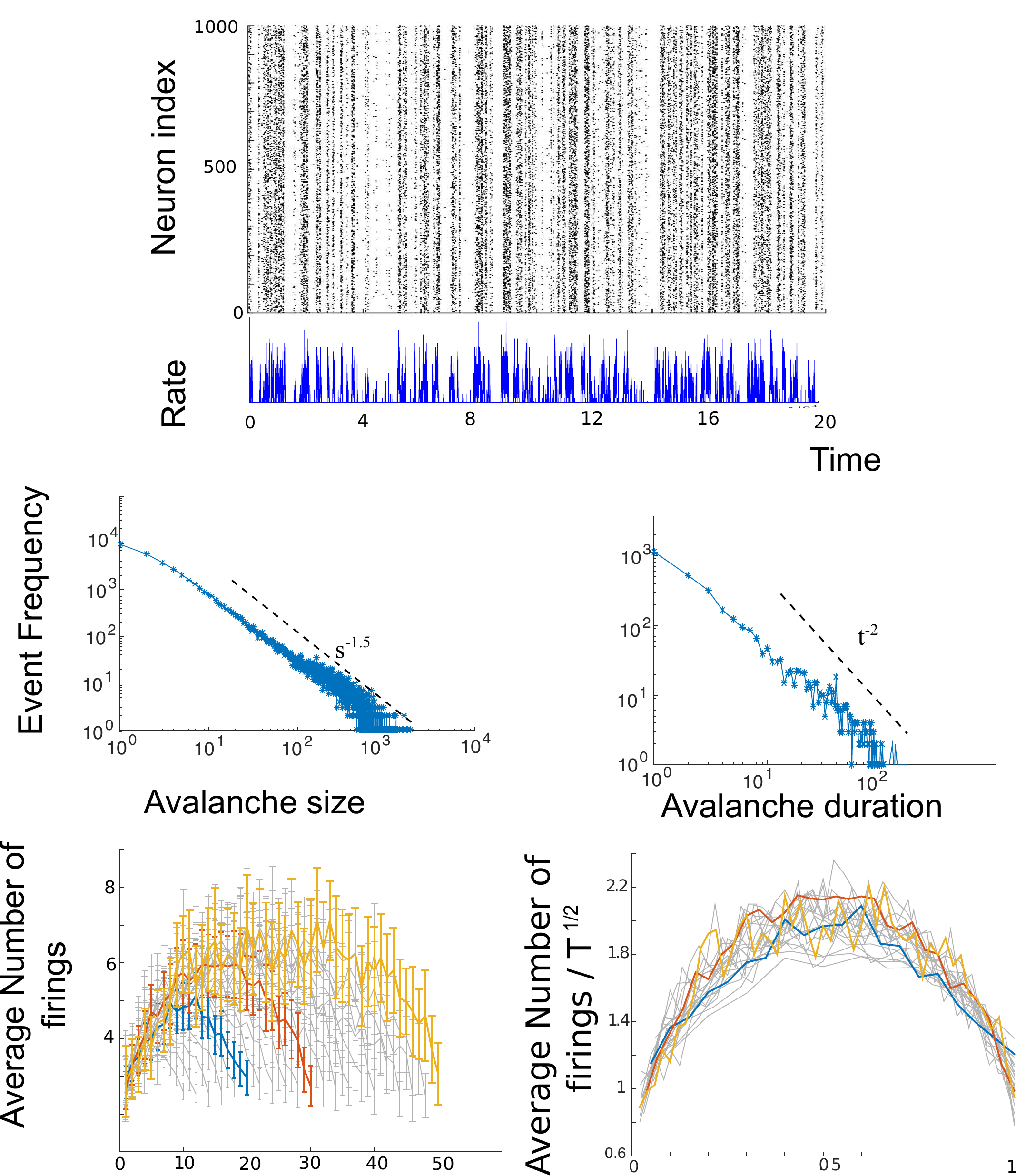}
	\caption{Avalanche statistics and shape collapses for independent Poisson processes with rates given by a reflected Brownian motion. }
	\label{fig:ReflectedBrownian}
\end{figure}
rates produce avalanches with power-law distributions of avalanche size and durations, consistent with our theory. To this purpose, we have performed a similar analysis as in Fig.~\ref{fig:Poisson} replacing Ornstein-Uhlenbeck firing rates by the positive part of a Brownian motion reflected at $\pm 1$. This choice was motivated by two constraints: the positive part was taken in order to consider only positive firing rates for consistency, and the reflection at $\pm 1$ was forced in order to prevent from having to long excursions of the Brownian motion, so that we can indeed assess that the heavy tails of the avalanche distributions are rather due to the statistical structure of the firings rather than due to possible very long excursions of the Brownian motion. The results of the simulations are provided in Fig.~\ref{fig:ReflectedBrownian}.  As in the case of the positive part of the Ornstein Uhlenbeck process, we find very clear power-law distributions of avalanche size and durations, with slopes consistent with our theory, and a very clear collapse of the avalanche shapes. 

\john{We add that beyond the collapse of the avalanche trajectories, the shapes onto these avalanche collapse may convey important information, as noted and investigated in the context of one-dimensional random walks~\cite{colaiori2004average,baldassarri2003average}. We observe indeed that the shape of the network-generated avalanches are not similar to the shapes obtained in the Brownian or Ornstein-Uhlenbeck case, and may similarly contain an information that goes beyond pure shape collapse reported in neural data~\cite{friedman2012universal}.}

\bibliographystyle{apsrev4-1}
\bibliography{../Bibliography}

\end{document}